\documentclass[12pt,a4paper]{article}
\usepackage{color,graphics,amsmath,epsfig,rotating}

\begin{document}

\newcommand{\ds}{\texttt{DarkSusy}}
\newcommand{\micro}{\texttt{micrOMEGAs}}
\newcommand{\py}{\texttt{PYTHIA}}
\newcommand{\HE}{\texttt{HERWIG}}
\newcommand{\msun}{{\rm M_{\odot}}}
\newcommand{\ie}{{\it i.e.} }
\newcommand{\eg}{{\it e.g.} }
\newcommand{\pbar}{${\rm \bar{p}}$}
\newcommand{\dbar}{${\rm \bar{D}}$}
\newcommand{\beq}{\begin{equation}}
\newcommand{\eeq}{\end{equation}}
\def\rsun{r_\odot}
\newcommand{\noi}{\noindent}

\newcommand{\calchep}{\texttt{CalcHEP}}

\begin{center}

{\Large\bf Indirect search for  dark matter with micrOMEGAs$\_ 2.4$} \\[8mm]

{\large   G.~B\'elanger$^1$, F.~Boudjema$^1$, P.~Brun$^2$,    A.~Pukhov$^3$, S.~Rosier-Lees$^4$,
P.~Salati$^1$, A.~Semenov$^5$. }\\[4mm]

{\it 1) LAPTH, Univ. de Savoie, CNRS, B.P.110,  F-74941 Annecy-le-Vieux, France\\
     2) CEA,Irfu, Service de Physique des Particules, Centre de Saclay, 
     F-91191 Gif-sur-Yvette, France\\
     3) Skobeltsyn Inst. of Nuclear Physics, Moscow State Univ., Moscow 119992, Russia\\
     4) LAPP, Univ. de Savoie, CNRS, B.P.110,  F-74941 Annecy-le-Vieux, France\\
     5) Joint Institute of Nuclear research, JINR, 141980 Dubna, Russia  }\\[4mm]

\end{center}

\begin{abstract}
We present a new module of \verb|micrOMEGAs| devoted to the computation of indirect signals from 
dark matter annihilation in any new model with a stable weakly interacting 
particle. The code provides the mass spectrum, 
cross-sections, relic density and exotic fluxes of gamma rays, positrons and antiprotons. 
The propagation of charged particles in the Galactic halo is handled with a new module that allows to 
easily modify the propagation parameters. 
\end{abstract}

\section{Introduction}

Cosmological observations show  strong  evidence that our Universe contains a large amount of dark matter (DM). 
New weakly interacting massive
particles (WIMP), such as those present in extensions of the Standard Model,  have roughly the correct annihilation 
properties to fit the high precision cosmological measurements.  Several astroparticle experiments are
actively searching directly or indirectly for this new particle. 
Indirect detection of dark matter particles involves  observation of the products of the DM annihilation in the galactic center, 
galactic halos or the extra galactic region. 
The annihilation products include positrons, anti-protons, anti-deuterons, gamma-rays and neutrinos.
Recently  many new results from indirect DM searches have been released.
Hints of  excesses that might be due to annihilation of dark 
matter particles have been reported although an interpretation of the measurements 
in terms of either DM annihilation or some astrophysical source  has not been confirmed.  
PAMELA shows an excess in the  positron fraction  between 10 and 100GeV~\cite{Adriani:2008zr}
in agreement with earlier indications by HEAT~\cite{Beatty:2004cy} and AMS01~\cite{Aguilar:2007yf}.  On the other hand PAMELA sees no
excess in the antiproton spectrum 
~\cite{Adriani:2008zq}. Both  Fermi~\cite{Abdo:2009zk}  and ATIC~\cite{Chang:2008zzr}  report an excess in the total electron 
plus positron spectrum  but at energies of several hundred GeV's, much above those of PAMELA. 
Furthermore the electron spectrum measured by HESS~\cite{Aharonian:2009ah} at very high energies is
consistent with both Fermi and ATIC. 
The cosmic gamma-rays from the galactic center or from galactic sources have been probed in a wide
energy range by INTEGRAL~\cite{Strong:2005zx}, Veritas~\cite{Maier:2008vw}, EGRET~\cite{Thompson:2008rw}, 
as well as HESS~\cite{Aharonian:2006au, Aharonian:2009nh} and Fermi~\cite{Meurer:2009ir}.
 These observations lead to  upper bounds on the DM 
annihilation cross section that are however strongly dependent on the halo profile, the propagation parameters and the background that is assumed for
the standard astrophysics processes. 
Observations in all channels are being pursued actively  with in
particular Fermi and  HESS taking data  as well as  AMS02 to be launched in 2011.  

The interpretation of the recent and upcoming data  requires  tools to compute
accurately the signals of DM annihilation in various channels  and this in the context of different particle physics models. 
The purpose of the package presented here is to compute indirect signals in $\gamma$, $e^+$ and \pbar~ produced in 
DM  annihilations in the Galaxy. 
This package is presented as a new module of \micro~\cite{micro13,Belanger:2008sj}, a code that computes the dark matter relic density,
the elastic scattering cross sections of WIMPs on nuclei relevant for direct detection as well as the cross sections 
and decay properties of new particles relevant for collider studies.\footnote{ DarkSUSY is another public code that computes the indirect signatures of
dark matter annihilation~\cite{darksusy}. It is confined to the minimal supersymmetric model.} 
\micro~ includes several models of particle physics that predict a new stable weakly interacting neutral particle,
the minimal supersymmetric standard model(MSSM) and several of its extensions, models with extra dimensions or the  little Higgs model 
as well as facilities to incorporate new models~\cite{Belanger:2006is}. 
As in earlier versions,  \micro~ provides the cross sections for dark matter annihilation into SM particles and the spectrum for 
$\gamma$, $e^+$ and \pbar~ at the source. The propagation of charged particles through the Galaxy which strongly distorts
 the charged particles 
spectra is the main addition  in this version. 

The main features included in the indirect detection module are: 

\begin{itemize}
\item
Annihilation cross sections for all 2-body tree-level processes for all models.
\item
Annihilation cross sections including radiative emission of a photon for all models.
\item
Annihilation cross sections into polarised gauge bosons.
\item
Annihilation cross sections for  the loop induced  processes  
$\gamma\gamma$ and $\gamma Z^0$ in the MSSM. 
\item 
Modelling of the DM halo with a general parameterization and with the possibility of taking into account DM clumps.
\item
Integrals along lines of sight for $\gamma$-ray signals.
\item
Computation of the propagation of charged particles through the Galaxy, including the possibility to modify
the propagation parameters.
\item
Effect of solar modulation on the charged particle spectrum.
\item
Model independent predictions of the  indirect detection signals. 
\end{itemize}

The neutrino spectrum originating from dark matter annihilation is also computed, however the neutrino signal is usually dominated 
by neutrinos coming
from DM capture in the Sun or the Earth. The inclusion of this signature is left for a further upgrade. 

In this paper we first review the procedure to obtain the flux of photons or anti-particles. This includes
the computation of  DM annihilation into SM particles and  a description of the dark matter halo models.  The 
propagation of charged particles including the issue of solar modulation is described in Section 4. 
The functions available in \micro~ are described in section 5 \footnote{\micro2.4 contains all the routines 
available in previous versions although the the format used to call some routines has
been changed. In particular global variables are used to specify the input parameters of various routines. 
All functions of \micro~ are described in the manual, manual4.tex, 
to be found in the main directory.}. Sample results as well as comparisons with other codes are
presented in Section 6.  

\section{Fluxes from DM annihilation}
\label{signals}

Should primordial self-annihilation take place in the early Universe, the same process would take place nowadays in the
denser regions of the Galactic DM halo. 
DM annihilation in the Galactic halo produces pairs of Standard Model particles that hadronize and decay into stable particles
These particles then evolve freely in the interstellar medium. 
The final states with  $\gamma$, $e^+$ and \pbar~ are particularly interesting as  they are the  subject of indirect 
searches.  The production rate of particles from DM annihilation at location {\bf x} reads
\beq
\label{eq:DMflux}
Q_a({\bf x},E)\;\;=\;\;\frac{1}{2} \langle\sigma v\rangle \left(\frac{\rho({\bf x})}{m_\chi}\right)^2 f_a(E)\;\;,
\eeq
where $\sigma v$ is the  annihilation cross-section 
times the relative velocity of incoming DM particles which we evaluate in the limit $v=0$ (this
is a good aproximation since $v=10^{-3}$). Note that $\langle\sigma v\rangle$ includes averaging over incoming
particles/antiparticles. For a Dirac particel where $\sigma_{\chi\chi}=0$, $\langle\sigma
v\rangle=1/2\sigma_{\chi\bar\chi}$.
 $m_\chi$ is the  mass of the DM candidate,  
$\rho({\bf x})$ is the DM mass density at the location ${\bf x}$ 
and  $f_a(E)=dN_a/dE$ is the energy distribution of the particle $a$ produced in one reaction.
The predictions for the energy spectra can also depend on some non perturbative effects including  
 QCD  and imply the use of Monte Carlo simulations.

\subsection{Annihilation cross-sections and energy spectra}
The cross-sections for the different 2-body
annihilation channels of WIMPs are calculated automatically in
any model implemented in \micro~\cite{Belanger:2006is}. This is done through the interface with 
\calchep~\cite{Pukhov:2004ca}.  All cross-sections are computed for 
a relative velocity $v=0$. There is an option in the code to define another value for $v$. 
The continuum spectrum for $\gamma$, $e^+$, $\bar{p}$, $\nu$ production is calculated as follows.

The self annihilation of DM particles can occur  at tree-level through 16 possible final states involving only pairs of SM particles. 
This includes all flavour diagonal pairs of fermions  
as well as gauge bosons, $\chi\chi\rightarrow q\bar{q},l^-l^+,\nu_l\bar{\nu}_l,W^+W^-,ZZ$.  
Other final states involving R-parity even particles are also included, in particular the Higgs. In the MSSM this corresponds to
all the channels with SUSY Higgs particles, namely $Zh_i,h_ih_j$,  $W^{\pm}H^{\mp}$ and $H^+H^-$ where $h_i$ stands for $h,H,A$. 
For the basic channels, $q\bar{q},l^-l^+,W^+W^-,ZZ$ and $gg$, we provide tables
for $\gamma$, $e^+$, $\bar{p},\nu$ production as obtained with PYTHIA version 6.4~\cite{pythia}.
 The database has been processed with $2\times 10^6$ events  at 18 fixed energies 
corresponding to $10<m_\chi< 5000$ GeV.
Note that neutrons and antineutrons do not decay in \py, basically a  $\rm \bar{n}$ is considered to be a \pbar~ and
the small amount of energy lost in the $\beta$ decay of $\rm \bar{n}$ is neglected. 
For channels containing two different
particles, $AB$, we obtain the final spectrum by taking half the sum of the $A\bar{A}$ and $B\bar{B}$ spectra.
For channels with Higgses, or other particles whose mass are a priori unknown,
we recursively calculate all $1\to 2$ decay channels until we
obtain particles in the basic channels. If during these decays we
get a pair of particles $AB$ where $A$ is one of the basic
channel, we suppose that $A$ gives half of the spectrum  obtained from
$A\bar{A}$  and we continue to decay $B$. 

The  $\gamma$ and $e^+$ spectra  can be substantially modified. 
First, polarisation of the gauge bosons final state can distort the positron and photon spectrum. 
Second, higher-order processes can also significantly modify the particle spectra. 
For example, photon radiation can strongly enhance some channels, this   
is particularly important for annihilation of a Majorana DM candidate into light fermions
which  suffers a s-wave suppression. The additional photon removes this suppression and the cross section increases 
by several orders of magnitude~\cite{Bergstrom:1989jr}. The implementation of these two effects in \micro~ are discussed in the following subsections. 
Finally at the one loop level, other 
final states are possible, such as $\gamma\gamma$, $\gamma Z^0$, $gg$  and $\gamma h^0$. Although supressed by a factor $\alpha^2$, the
processes with $\gamma$'s are nevertheless interesting since they lead to a spectacular signal, a mono-energetic $\gamma$ ray line.
The one-loop processes $\chi\chi\rightarrow \gamma\gamma,\gamma Z^0$ have been  computed automatically with 
{\bf Sloops} for the  MSSM~\cite{Boudjema:2005hb,Baro:2008bg}, and are incorporated in ~\micro. Note that the Majorana nature of the neutralino
forbids an annihilation into   $\gamma h^0$ at rest.

\subsubsection{Vector boson polarisation}

The primary particles produced in DM annihilation are by default assumed to be unpolarised. 
In general however these particles and in particular vector particles can be polarised. 
For example the annihilation of neutralinos in the MSSM produces only
transverse W's and Z's while the polarisation of spinor particles can be neglected because of the CP invariance of
the initial state. The spectra after decay and hadronisation of standard particles  extracted from
PYTHIA also assumes unpolorized particles. 

To take into account the polarisation, 
we include an option for gauge bosons pair production. The first step is to determine the 
degree of polarisation  of the vector bosons produced via dark matter anihilation in a given
model. We only need to determine the polarisation of one of the vector bosons as only the
$V_TV_T$ or $V_LV_L$ combinations are possible. To do this automatically, we compute  with \micro~  the three-body process $ \chi\chi\to W^-e^+\nu_{e}$
keeping only the contribution from W pair production followed with the $W^+$ leptonic decay. 
We then check numerically the angular distribution  of $\nu_{e}$
in the rest frame  of the  on-shell $W^+$ with respect to the direction of flight of the W pair. 
The angular distributions are expected to be
$ 3/8(1 + \cos^2\theta) d\cos\theta$ for $W_T$ and  
$ 3/4(1-\cos^2\theta)d\cos\theta$ for $W_L$. 
With this method, we can reconstruct automatically the 
 $W^+$ polarisation  in a generic model.

We then need the spectra of the stable particles produced after decay and hadronization of a polarised gauge boson. 
For this, two methods have been used
and compared showing  perfect agreement.  In the first method, we pass to PYTHIA 
events with four outgoing particles representing the decays
$W^+\to 2x$ and  $W^-\to 2x$ where the decay products 
are distributed according to the formulas presented
above. 
For this  we used PYTHIA 6.4 and the Les Houches event interface \cite{Boos:2001cv}.
Initial events  were generated  only for $M_\chi=M_W$. Results for heavy CDM 
where obtained by boosting. The same procedure is also used for neutral gauge bosons.
In the second method, a reweighting technique is applied within PYTHIA 6.4, by measuring event by event the $\theta$ angular distribution 
of the primary W (or Z) decay fermions in the boson rest frame.  
Namely, for each primary fermion a weight  is determined  depending on the polarisation assumption. 
For longitudinal polarisation it is equal to $\frac{3}{4}\times(1-\cos^2\theta)$ while
it becomes  $\frac{3}{8}\times(1+\cos^2\theta)$ for transverse polarisation as explained previously.
Then for each stable particle ($\gamma, e^+ \bar{p},\nu$), the weight of the W ( or Z) decay fermion they originate from,
is used to build the energy spectrum for each polarisation scenario.  This is done for 18 different CDM masses. 
Extensive comparisons of spectrum distributions for direct or reweighted longitudinally polarized bosons have been made. 
They are in perfect agreement.  We also checked that the average of the polarised spectra obtained with each method agreed 
with the unpolarised one. 

\begin{figure}[h]
\centering
\includegraphics[width=12cm]{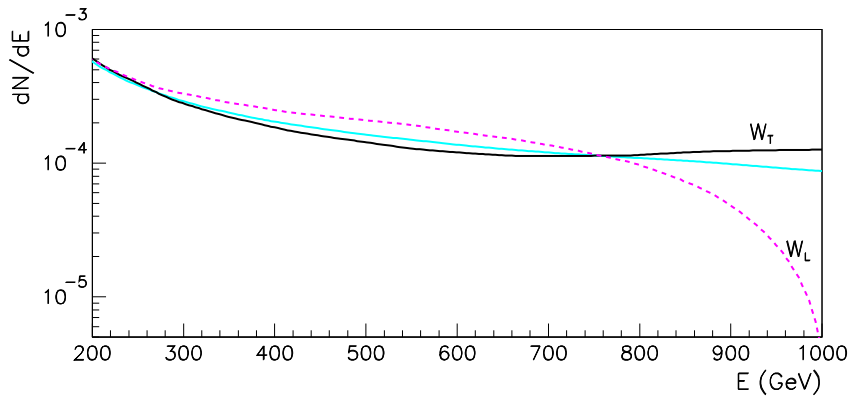}
\caption{\label{comparison_Wpol} $dN/dE$ for positrons from DM  annihilating into $W^+W^-$  with transversely polarised (full/black), 
longitudinally polarised (dash/pink) or unpolarised W's (full/blue),  here $m_\chi=1$~TeV. }
\label{fig:wpolar}
\end{figure}

Taking into account the polarisation of the W's (or Z's) leads to a harder spectrum for $e^+$ originating from transverse W's than if the W's were
assumed unpolarised. The polarisation effect 
corresponds to at most a factor of 3/2 increase for the most energetic charged particles.  This is because 
 both transversely polarised W give a harder spectrum while the spectrum for longitudinally polarised W vanishes at high energy, 
 see fig.~\ref{fig:wpolar} for the positron spectrum.
 Both the polarised and unpolarised spectra are available in the {\tt basicSpectra} routine.
To calculate the spectra of DM annilation taking into account the polarisation of  vector bosons
one has to set a switch in the {\tt calcSpectrum} routine, see the routines description in Section~\ref{code}.

\subsubsection{Photon radiation}

In a DM annihilation process, photons can be emitted from the external legs,  final state radiation  (FSR),
from an internal leg, from a quartic vertex  or in the subsequent decay of the pair of particles directly produced.  
The first two processes cannot be treated separately in a gauge invariant manner and
are associated with a three-body annihilation process.  They  are a priori suppressed by a factor $\alpha$, 
however large enhancements can occur~\cite{Bringmann:2007nk}.  In particular final state radiation features a collinear log corresponding to a photon emission from an external leg. 
This virtuality  gives a factor  $\log(4 M_{\chi}^2/m_f^2) $ where $m_f$ is the mass of the outgoing particle. 
For light particles final state radiation can therefore  be approximated by the production cross section for two particles times a radiation factor that only depends on the
spin of the particle. This is the approach followed in PYTHIA.  
Furthermore in the case of light masses $m_f \ll M_{\chi}$ one expects multiple photon radiation for the soft spectrum, this is also used in PYTHIA
so that the FSR spectrum differs from the one corresponding to a single photon radiated from an external leg by up to 10\% - 20\% in the high energy 
 part.
For W bosons the virtuality is never large so photon radiation is ignored in PYTHIA, the only photons included in PYTHIA are those coming from the decays of the
gauge bosons. In this case a full $2\rightarrow 3$ calculation is in order, 
however barring exceptional cases the yield is suppressed by a factor
$\alpha$ (with no large logarithm enhancement) and therefore
small. An important example where the approximation, and
therefore applying PYTHIA, fails is if the cross section
without radiation of a photon is for symmetry reasons very small
or vanishing so that factorisation  does not hold. A notable case, mentionned above,  is the
annihilation of the Majorana neutralino, at $v=0$, to a pair of
almost massless fermions.
The $s$-wave cross section of order the chirality factor
$m_f^2/M_{\chi}^2$ can be much smaller than the radiation cross
section of order $\alpha$ since the chirality argument no longer applies once a
photon is radiated. 
The cross section including a photon emission can 
be enhanced in some  specific cases. This can occur in situations where the $t$-channel
particle exchanged between the DM particle and the
external charged particle is not far from the DM mass, leading to
an  enhancement factor ${ M_{\chi}^2}/(M_{I}^2 - M_{\chi}^2)  $.
$M_I$ is the mass of the internal particle.

In our code we compute the direct photon production through the full $2 \to
3$ calculation, those are generated at
run-time but only for the situation of interest which we have
defined as $M_I^2 < 1.5 M_{\chi}^2$. We also compute the full 
$WW\gamma$ in situations where 
$M_\chi > 500$~GeV because of the potential large log enhancement due to photon emission since in this situation the W can be relativistic. 
 All 3-body final states are included in
the computation of the photon spectrum when this option is chosen.
On the other hand only the $e^+e^-\gamma$ 
process is included in the positron spectrum. Indeed these are
the only processes that affect significantly the hard part of the
positron spectrum. 

When the full $\chi \bar{\chi} \to f \bar f \gamma$ process is
generated by the code, one should avoid double counting due to the
inclusion of the photons induced by PYTHIA  through the direct  $f \bar f
\gamma$. By direct we mean the photons generated prior to the
possible decays of $f$ and hadronisation.  
To subtract the FSR contribution already taken into account in PYTHIA 
we remark that our generated $\chi \bar{\chi} \to f \bar f
\gamma$ photon must exhibit the infrared divergent behaviour
$1/x$, if the $\chi \bar{\chi} \to f \bar f $ cross section is not
vanishing. This infrared behaviour at $x=0$ can only originate
from radiation from external charged legs.
We therefore
expand the generated $2 \to 3$ cross section of our code
${d\sigma v}/{dx}$ around $x=0$ and write
\begin{equation}
\label{expand2to3} \frac{d\sigma v}{dx} = \frac{A}{x}+ B + C x
\end{equation}
The idea is to subtract from our generated cross section the $A/x$
term obtained for small enough $x$. For this  we use $x=0.01$. We have
checked that $0.001<x<0.03$ give similar results. There might still be
a mismatch between the coefficient $A$ extracted from the full
calculation and the one contained in PYTHIA. This difference is
due to QCD hadronisation, higher order terms, additional photons and choice of the
scale $Q^2$ for the splitting, contained in PYTHIA. To conform
with the splitting function, for light fermions for example, we
coud then subtract $A (1-x+x^2/2)/x$,
however in our code this hardly makes a difference in situations
where the full calculation is important.
At the level of implementation let us mention
that there might be a problem caused by the finite precision of
the phase space integration of the  $2\to3$ matrix elements. For a
fixed photon energy, there is an infrared pole when the outgoing
fermion and the radiated photon become collinear. Furthermore
strong numerical cancellations occur when summing over all diagrams.
In the case of small fermion masses this can lead to numerical
instability. To solve this we replace the integration variable
$d\cos{\phi}$ to $d\phi$ ($\phi$ is the angle between the outgoing
fermion and the radiated photon in the rest frame of the fermion pair). This trick works well for $m_f
< 10^{-4}m_{\chi}$. When $m_f< 10^{-3}m_{\chi}$ we keep
$d\cos{\phi}$ as the integration variable and  impose a cut 
$|\cos\phi| < 0.99$, in this region the numerical calculation is  robust. The poles at
$\phi=m_f/M_{\chi}$ due to radiation from the final charged
particles are safely out of the integration region. We have tested
that both methods are in good agreement  when   $m_f =
10^{-3}m_{\chi}$.

\begin{figure}[h]
\centering
\includegraphics[width=.52\columnwidth]{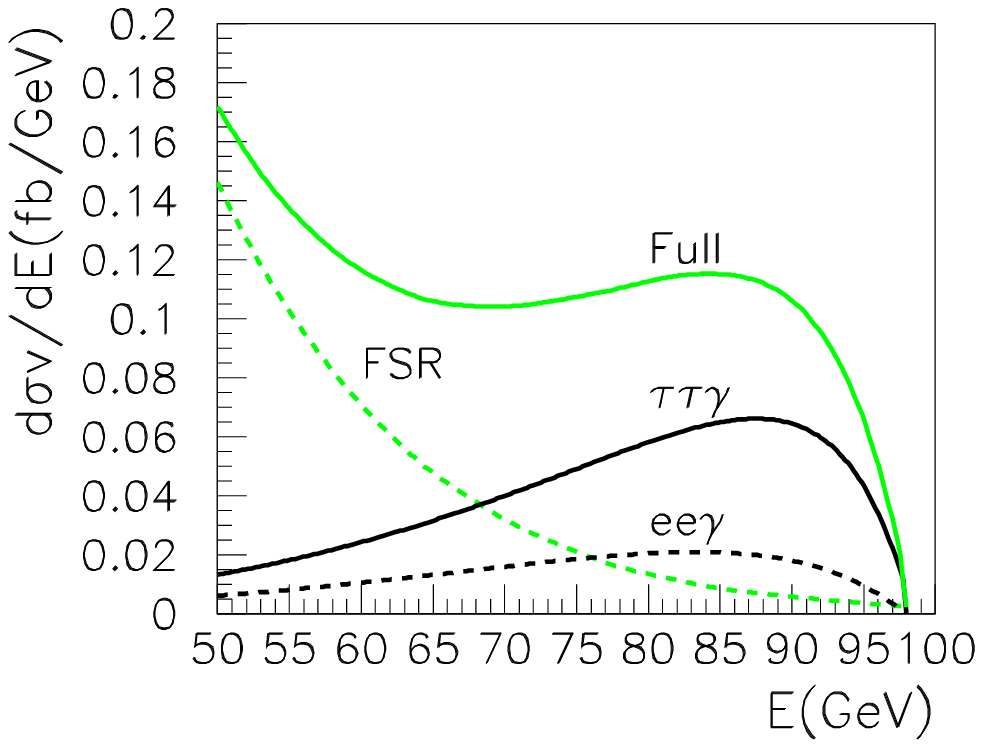}
\includegraphics[width=.45\columnwidth]{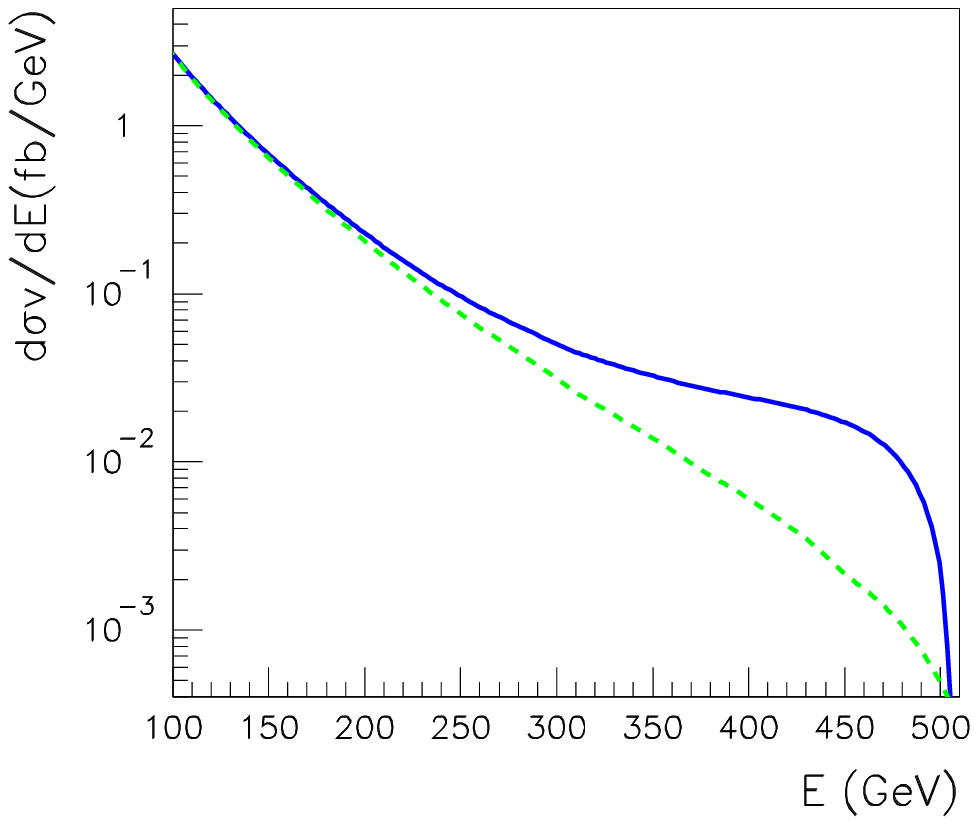}
\vspace{-.8cm}
\caption{\label{fig:photon} a) Photon spectrum for the CMSSM point $m_0=70$~GeV, $M_{\frac{1}{2}}=250$~GeV,
$A_0=-300$~GeV, $\tan\beta=10$ for $\chi\chi\rightarrow \tau^+\tau^-\gamma$ (black), $e^+e^-\gamma$
(black-dash), with photons from external legs from  PYTHIA  (FSR) (green-dash) and all  contributions from 3-body final states (green-full). 
b) Photon spectrum for a MSSM point ($\mu=545 {\rm GeV}, M_1=500 {\rm GeV}, \tan\beta=20, M_A=2{\rm TeV}$ and
 $m_{\tilde f}=2.5{\rm TeV}$) with (full) and without (dash) the contributions from the three-body process
$\chi\chi\rightarrow W^+W^-\gamma$}
\end{figure}

The importance of the extra contributions not contained in the
factorised FSR to the photon spectra is illustrated in Fig.~\ref{fig:photon}a
for the CMSSM point $m_0=70$~GeV, $M_{\frac{1}{2}}=250$~GeV,
$A_0=-300$~GeV, $\tan\beta=10$. Here $M_{\chi}=97.9$~GeV,
$M_{\tilde{\tau}}=107.6$~GeV,
$M_{\tilde{e}}=M_{\tilde{\mu}}=123.9$~GeV, neutralinos annihilate dominantly into tau pairs, this
is an example of a case where the t-channel particle exchanged, the $\tilde\tau$,    
is not far from resonance so that photon radiation from internal lines is enhanced.  
Note that  the curve  denoted $\tau\tau\gamma$ in Fig.~\ref{fig:photon}a represents only the contribution from the 3-body process,  
the additional photons that originate from $\tau$ decays are included  when computing the full process.  
The enhancement in the high-energy
part of the photon spectrum in a case where DM  annihilation into gauge bosons pairs is dominant 
is illustrated in Fig.~\ref{fig:photon}b for the MSSM. Here the DM is a mixed bino/higgsino LSP with
 $\mu=545 {\rm GeV}, M_1=500 {\rm GeV}$, $M_3=6M_1=3M_2$, $\tan\beta=20, M_A=2{\rm TeV}$ and
all soft sfermion masses heavy, $m_{\tilde f}=2.5{\rm TeV}$.

\subsection{Dark matter halo models}

In \micro~ routines which calculate the propagation of particles in the
Galaxy the DM halo distribution is an input parameter. Thus
\micro~ can work with any sphericaly symmetric DM halo
profile. As an example of DM halo distribution  we  include 
a widely  used spherically symmetric parametrization of the dark matter halo  
\begin{eqnarray}
\rho(r)&=&\rho_{\odot} F_{halo}(r) \nonumber\\
F_{halo}(r) &=& \left [ \frac{r_{\odot}}{r} \right ] ^{\gamma}
\left [ \frac{1+(r_{\odot}/a)^{\alpha}}{1+(r/a)^{\alpha}} \right ]
^{\frac{\beta-\gamma}{\alpha}}
\label{halo}
\end{eqnarray}
The values for the $\alpha$, $\beta$, $\gamma$ and $a$ parameters for the most common halo models are listed in
 Tab.~\ref{tab:halo}, the default values are those of the NFW profile.  
The value of the  DM density at the solar location $\rho_{\odot}$ 
and $r_{\odot}$  the distance of the Sun to the Galactic center  are global  parameters of \micro. Their value can be
redefined simply and the default values are listed in Table 3. 
An important remark concerns the central divergence. For $\gamma \geq 1.5$ in eq.~\ref{halo} there is an non integrable 
squared density in the center of the galaxy.
To avoid this singularity we set a limit for the distance from the Galactic center
 $r > r_{min}=0.001pc$  in our integration routines. This is done for any halo profile.

\begin{table}[h]
  \centering
    \begin{tabular}{c|c|c|c|c}
    Halo model & $\alpha$ & $\beta$ & $\gamma$ & a (kpc) \\
    \hline
    Isothermal with core & 2 & 2 & 0 & 4 \\
    \hline
    NFW & 1 & 3 & 1 & 20 \\
    \hline
    Moore  & 1.5 & 3 & 1.5 & 28 \\
    \end{tabular}
    \caption{Halo parameters for three common profiles}
    \label{tab:halo}
\end{table}

The user also has the possibility to  use a totally different  parameterization for the halo, provided it is spherically 
symmetric. For example the Einasto profile has been advocated recently~\cite{einasto}
\beq
 F_{halo}(r)= \exp\left[ \frac{-2}{\alpha}\left( \left(\frac{r}{r_{\odot}}\right)^\alpha-1\right) \right]
\label{einasto}
\eeq
where the default value for $\alpha$ is $0.17$.

DM annihilation depends on the squared density  $\overline{\rho^2}(r)$.
In general a clumpy structure of DM will lead to  $\overline{\rho^2}(r)  \;> \; \overline{\rho}(r)^2$.
The effect of clumping can be described by another profile 
\begin{equation}
\overline{\rho^2}(r)=\rho^2_\odot F_{halo}^2(r)F_{clump}(r)
\end{equation}
The presence of clumps  will lead to an enhancement factor or boost factor.
In general the  clump number density in a volume bounded by the characteristic diffusion length of the involved
species will determine the size of the enhancement factor. The resulting boost factor can  therefore differ for 
photons, positrons or antiprotons. In the last two cases it does not typically exceed 20~\cite{Lavalle:1990wn} . 
Larger boost factors can be found for some extreme
clump configurations, for example a big sub-halo close to the Earth~\cite{Lavalle:1990wn}, although this situation is very unlikely
~\cite{Hooper:2003ad}.

\subsection{Photons}

The gamma ray flux can be evaluated as
\begin{equation}
   \Phi_\gamma(E,\phi) = \frac{\sigma v}{m^2_\chi} f_\gamma(E) H(\phi)\;\; \label{flux}
\end{equation}
and is expressed in number of photons per $cm^2 s \; sr$. The factor $H$ includes the integral of the squared of the dark
matter density over the line of sight,
\begin{equation}
H(\phi) = \frac{1}{8\pi}\int_0^{\infty} dr
\overline{\rho^2}(r')
\label{halo_photon}
\end{equation}
where $r'=\sqrt{r^2+r_\odot^2-2rr_\odot\cos\phi}$ and $\phi$ is the angle in the direction of 
observation. In \micro~ one can compute the photon flux performing the integral over the line of sight and over the opening angle
 which characterizes the detector resolution, see section~\ref{code}.

\section{Galactic propagation of charged particles}

\subsection{General framework}
The charged particles generated from DM annihilation propagate through the Galactic halo and their energy spectrum at the Earth differs
from the one produced at the source. 
Charged particles are deflected by the irregularities of the galactic magnetic field. 
In the Milky Way which has strong magnetic turbulence Monte
Carlo simulations indicate that this is described by  space diffusion. 
Charged particles also suffer energy losses from synchroton radiation and inverse Compton scattering as well 
as diffusive reacceleration in the disk.  Finally
galactic convection wipes away charged particles from the disk. 
Solar modulation can also affect  the low energy part of the spectrum. The  equation that describes the evolution of the 
energy distribution for all particles (protons, anti-protons, positrons) reads
\begin{equation}
\label{eq:propa}
\frac{\partial}{\partial z} \left(V_C\psi_a\right)
- {\bf \nabla}\cdot\left( K(E) {\bf \nabla} \psi_a \right)-
\frac{\partial}{\partial E} \left( b(E) \psi_a \right) =Q_a({\bf x},E)
\end{equation}
where $\psi_a=dn/dE$ is the number density of particles per unit volume and energy, $a$ denotes the particle specie and
$Q_a$ is the source term. Here we do not consider background and  include only  particles produced from dark matter
annihilation, Eq.~\ref{eq:DMflux}. For antiprotons a negative contribution to the source term will also be considered to account
for antiproton annihilations in the interstellar medium
(see section~\ref{sec:pbar}).
$K$ is the space diffusion coefficient, assumed homogeneous. 
\begin{equation}
K(E)=K_0 \beta(E) \left({\cal R}/1\;{\rm GV}\right)^\delta
\end{equation}
where $\beta$ is the particle velocity and ${\cal R}=p/q$ its rigidity.
 $b(E)$ is the energy loss rate. Another term that describes the diffusive reacceleration has been neglected. 
 The simple power law for  $K(E)$ is inferred from magnetohydrodynamics considerations ~\cite{Ptuskin}, 
 once the convective velocity is taken into account it has been shown that this form for $K(E)$ was adequate 
 to fit the B/C data~\cite{Maurin:2001sj}.

The propagation equation, Eq. ~\ref{eq:propa}, is solved within a semi-analytical two-zone model which was discussed in 
~\cite{Berezinskii,Maurin:2001sj, Donato:2001ms}.
Within this approach the region of diffusion of cosmic rays is represented by a thick disk 
of thickness $2L$ and radius $R\approx20$~kpc. The thin galactic disk 
 lies in the middle 
and has a thickness $2h\approx 200$~pc and radius $R$. The boundary conditions are such that the number density vanishes at $z=\pm L$ and
at $r=R$.
The galactic wind is directed outward along the $z$ direction
so the convective velocity is also vertical and of constant magnitude $V_C(z)=V_C sign(z)$.
The propagation parameters $\delta, K_0,L,V_C$  are constrained by the analysis of the boron to carbon ratio,  
a quantity sensitive to cosmic ray transport~\cite{Maurin:2001sj}. 
Typical values for these coefficients are listed in Table~\ref{tab:BC}, default values are those of the MED model.

\begin{table}[h]
  \centering
    \begin{tabular}{c|c|c|c|c}
 Model & $\delta$ & $K_0\;(\rm kpc^2/Myr)$ & $L\;(\rm kpc)$ & $V_C(\rm{km/s})$ \\\hline
  MIN & 0.85 & 0.0016 & 1 & 13.5 \\\hline
  MED & 0.7 & 0.0112 & 4 & 12 \\\hline
  MAX  & 0.46 & 0.0765 & 15 & 5 \\
    \end{tabular}
    \caption{Typical diffusion parameters that are compatible with the B/C analysis~\cite{Maurin:2001sj,Donato:2003xg}}
    \label{tab:BC}
\end{table}

The solution for the energy distribution, eq.~\ref{eq:propa}, will generally be expressed as an integral equation
\begin{equation}
\psi (E,r_{\odot}) = 
\int_E^{m_\chi} dE_S \int d^3{\bf x}_S G({\bf x}_{\odot},E;{\bf x}_S,E_S) Q({\bf x}_S,E_S)
\end{equation}
where the last integral is performed over the diffusive halo.
$E_S$ is the energy at the source and $G({\bf x}_{\odot},E;{\bf x}_S,E_S)$ is the Green function 
which determines the probability for a cosmic ray
produced at ${\bf x}_S$ with energy $E_S$ to reach a detector at the Earth with an energy $E$. 
The differential flux is related to the number density
\begin{equation}
\Phi_a=\frac{v(E)}{4\pi} \psi_a
\end{equation}
where $v$ is the particle velocity. 
The method of solution of the general equation adapted to each type of charged cosmic rays will be described next.

\subsection{Positrons}
The energy spectrum of positrons is obtained by solving  the diffusion-loss
equation keeping only the two dominant contributions:  space diffusion and energy losses.
 \begin{equation}
- {\bf \nabla}\cdot\left( K(E) {\bf \nabla} \psi_{e^+} \right)-
\frac{\partial}{\partial E} \left( b(E) \psi_{e^+}\right) =Q_{e^+}({\bf x},E)
\end{equation}
Here $K=K_0 (E/E_0)^\delta$ since for energies above 
0.1~GeV the positrons are ultra relativistic and the rigidity ${\cal R}$ is proportionnal to $E$, $E_0=1GeV$.  
The positron loss rate  is dominated by synchrothon radiation in the galactic magnetic field and inverse Compton scattering on
stellar light and CMB photons, 
\begin{equation}
b(E)=\frac{E^2}{E_0\tau_E}
\end{equation}
where $\tau_E=10^{16}$~s is the typical energy loss time. Subdominant contributions from diffusive reacceleration and
galactic convection were shown to have an impact only below a few GeV's ~\cite{Delahaye:2008ua}, these effects are not
included in our treatment.

The propagation equation can be transformed into the  heat conductivity equation after the 
substitutions
\begin{eqnarray}
\psi(E,r,z)&=&\bar{N}(t,r,z)/b(E)\\
t(E)&=& - \int dE\frac{K(E)}{b(E)}
\end{eqnarray}
The propagation equation now reads 
\begin{equation}
\label{positron_eq_imp}
( \frac{\partial}{\partial t} -\nabla^2 )
\bar{N}(t,r,z) = \frac{\sigma v}{2 m_\chi^2} \overline{\rho^2}(r,z)
\left(\frac{f_{e^+}(E)b(E)}{K(E)}\right){\Bigg\vert}_{E=E(t)}
\end{equation}
Assuming $\overline{\rho^2}(r,z)=\overline{\rho^2(r,-z)}$,
we can solve Eq.~\ref{positron_eq_imp}
in the region $0<z<L$ with the  boundary conditions 
\begin{equation}
  \bar{N}(t,r,L)=0; ~~~~~~ \frac{\partial \bar{N}(t,r,z)}{\partial z}{\Bigg\vert}_{z=0}
=0
\end{equation}
The Green function for Eq.~\ref{positron_eq_imp} can be factored into a r-dependent part, a Gauss function, and
a $z$-dependent part. The latter is 
more complicated because of boundary conditions,
\begin{eqnarray} 
\label{positron_gf_1}
( \frac{\partial}{\partial \tau} - \nabla^2)
\bar{G}(\tau,r,z,z{'}) &=& \delta(\tau)\delta^2(\vec{r})\delta(z-z{'})\\
\label{positron_gf_2}
\bar{G}(\tau,r,z,z{'})&=& \frac{\Theta(\tau)}{4\pi
\tau}exp(\frac{-r^2}{4\tau}) g_v(\tau,z,z{'})\\
\label{positron_gf_3}
( \frac{\partial}{\partial \tau}  -\frac{\partial^2}{\partial z^2}
)g_v(\tau,z,z{'})&=&\delta(\tau)\delta(z-z{'})
\end{eqnarray}
To calculate $g_v$, the vertical Green function,  we construct a basis of eigenstates $e_n(z)$ of the operator
$-\frac{d^2}{dz^2}$ defined on an interval $0<z<L$ with the boundary conditions $e_n(L)=0$ and $e'_n(0)=0$. This basis is
\begin{equation}
e_n(z)= \sin(\pi(n+0.5)/L)
\end{equation}
The  vertical Green  function expressed in terms of $e_n(z)$ reads
\begin{equation}
\label{positron_green_v_basis}
     g_v(t,z,z{'})= \Theta(t)\sum\limits_{n=0}\limits^{\infty}
e^{-k_n^2 t} e_n(z)e_n(z{'})/c_n
\end{equation}
where $c_n$ are the normalization constants
\begin{equation}
\label{cn_norm}
    \int^L_0 e_n(z)e_m(z)dz=\delta_{nm}c_n
\end{equation}
When   $t$ is  small, eq.~\ref{positron_green_v_basis} does not  converge well. In this case 
$g_v$ is rather obtained using the method of electrical images,
\begin{equation}
    g_v(t,z,z{'})=\sum\limits_{n=-\infty}\limits^{\infty}
\frac{(-1)^{n}}{\sqrt{4\pi\tau}} \left(
  e^{-\frac{(z-z{'}+2nL)^2}{4\tau}} +  
  e^{-\frac{(z+z{'}+2nL)^2}{4\tau}}\right)
\end{equation}
Using the Green functions, Eq.~\ref{positron_gf_1} - \ref{positron_gf_3}, the  
positron density near the Sun is obtained after a $3D$ integration
\begin{eqnarray}
\psi_{\bar{e}}(E_0,r_{\odot},0)&=& \frac{\sigma v}{b(E_0)}
\int\limits^{M_\chi}\limits_{E_0} dE f(E)
D(t(E_0)-t(E),r_{\odot})\\
D(0,r_{\odot})&=&\frac{1}{2M_\chi^2}\overline{\rho^2(r_{\odot},0)}\\
\label{positrons_R_limit}
D(\tau, r_\odot )&=&\frac{1}{4 \tau} \int\limits_0\limits^L dz g_v(\tau,z,0)\int\limits_0\limits^{\infty}
rdr \frac{\overline{\rho^2}(r,z)}{M_\chi^2}
\exp\left({-\frac{(r-r_{\odot})^2}{4\tau}}\right)
I_{\phi}\left(\frac{r_{\odot}r}{2\tau}\right)\\
I_{\phi}(x)&=& \frac{1}{\pi} \int\limits^{\pi}\limits_0 d\phi
e^{-2x\sin^2\frac{\phi}{2}} = I_0(x)e^{-x} 
\end{eqnarray}
where $I_0$ is the modified Bessel function of the  first
kind. In practice we use $R$  as the upper limit of the integral over $r$ in  Eq.~\ref{positrons_R_limit}.
A more precise  treatment of the boundary condition is not required as the  
positrons originating  from far away sources suffer significant energy losses.
 
Note that $D(\tau,r_{\odot})$ is an universal function for all energies.
To calculate $\psi_{\bar{e}}(E)$ for all positron energies, it is more efficient 
to first tabulate $D$ as a function of $\tau$ in the 
region $0\le\tau\le t(E_{min}) -t(M_{\chi})$ and
then perform a fast integration for all energies. This is the method implemented in 
the \micro~  routine \verb|posiFluxTab|.

\subsection{Antiprotons propagation}
\label{sec:pbar}
The propagation of antiprotons is dominated by diffusion and the effect of the galactic wind. The source term includes the
annihilation of DM, Eq.~\ref{eq:DMflux}, as well as a negative source term corresponding to the annihilation of antiprotons in
the interstellar medium ($H,He$). The annihilation rate
\begin{equation}
\Gamma_{tot}=\sigma_{\bar{p}H}^{ann} v_{\bar{p}}n_H + \sigma_{\bar{p}He}^{ann} v_{\bar{p}}n_{He}
\end{equation}
where $v_{\bar p}$ is the velocity of the $\bar{p}$.
The annihilation cross-sections $\sigma_{\bar{p}H}^{ann}$ are found in \cite{Tan:1982nc,Tan:1984ha} and rescaled by a factor
$4^{2/3}$ for $\sigma_{\bar{p}He}^{ann}$. The average densities in the galactic disc are set to $n_H=0.9{\rm cm}^{-3}$ and 
$n_{He}=0.1{\rm cm}^{-3}$.
The production of secondary antiprotons is not included in the code.

The energy spectrum of antiprotons is obtained by solving the diffusion equation 
\begin{equation}
\label{aproton_eq}
\left[ -K(E)\nabla^2+V_c\frac{\partial}{\partial z}
+ 2(V_c  + h \Gamma_{tot}(E))\delta(z)\right]
\psi_{\bar{p}}(E,r,z)=\frac{\sigma v}{2} \frac{\overline{\rho^2}(r,z)}{M_\chi^2} f_{\bar{p}}(E)
\end{equation}
An important difference with the positron case is that energy loss of antiprotons is negligible, 
we will therefore omit the $E$ dependence  until the final formula.
 $\partial \psi(r,z)/\partial z $ should have a  discontinuity at $z=0$.
Assuming that $\rho^2(r,z)=\rho^2(r,-z)$  means that $\psi_{\bar{p}}(r,z)$ has the same 
symmetry and the discontinuity  can be presented  as a boundary condition
\begin{equation}
   \frac{\partial \psi_{\bar{p}}(r,z)}{\partial z}{\Bigg\vert}_{z=0} = \psi_{\bar{p}}(r,0)(V_c+h\Gamma_{tot})/K
\end{equation}
After substituting 
\begin{equation}
       \psi_{\bar{p}}(r,z)= \exp(k_c z) \bar{N}(r,z);
\end{equation} 
where $k_c=V_c/(2K)$,  Eq.~\ref{aproton_eq} simplifies to
\begin{equation}
\label{aproton_eq_imp}
\left[ -\nabla^2 +k_c^2
\right]\bar{N}(r,z)= e^{-k_c z}\frac{\sigma v}{2K}\frac{\overline{\rho^2(r,z)}}{M_\chi^2}
f_{\bar{p}}  
\end{equation}
for  $0<z<L$  with the boundary conditions 
\begin{equation}
\bar{N}(r,L)=0;~~~~~~~~~ \frac{\partial \bar{N}(r,z)}{\partial z}{\Bigg\vert}_{z=0} =
\bar{N}(r,0)v_d;  
\end{equation}
where $v_d= (V_c/2+h\Gamma_{tot})/K$. 

To compute the  Green function for Eq.~(\ref{aproton_eq_imp}) we proceed as for the positron case and   construct a basis of eigenstates 
of the $-\frac{d^2}{d z^2}$ operator defined on the interval $ 0<z<L$ 
with the boundary conditions
\begin{equation}
    e_n(z)=\sin{(k_n(L-z))}   
\end{equation}
where the  set of $k_n$ is defined by  the condition 
\begin{equation}
   e'_n(0)=e_n(0)v_d.
\end{equation}
The Green function then reads
\begin{equation}
\label{aproton_green}
 \bar{G}(r,z,z{'}) =\frac{1}{2\pi} \sum\limits^{\infty}_{n=0}
K_0(r\sqrt{k_c^2+k_n^2})e_n(z)e_n(z')/c_n
\end{equation}
where $c_n$ are the  normalization constants (see Eq.~\ref{cn_norm})
and $K_0$ is the MacDonald function defined by the equation
\begin{equation}
  \Delta_r K_0(r)-K_0(r)=-2\pi \delta^2(\vec{r})
\end{equation}

The antiproton energy spectrum at the Earth is obtained after a 
 $3D$ integration
\begin{equation}
\label{antiproton_answer}
   \psi_{\bar{p}}(E,r_{\odot},0) = \frac{\sigma v f_{\bar{p}}(E)}{K}\int^{\infty}_0
r dr \int^L_0 dz  \bar{G}(r,z,0)e^{-k_c z} \int^{\pi}_0 d\phi
\frac{\overline{\rho^2(r,z')}}{M_\chi^2}
\end{equation}
with $ r'=\sqrt{r_{\odot}^2+r^2+2r_{\odot}r\cos{\phi}}$.

To avoid numerical problems  due to a possible singularity in the dark matter density 
near the center of the Galaxy, we integrate Eq.~\ref{antiproton_answer}
in the central region  $|x|<r_0=0.01kpc$ fixing   $\rho^2(|x|)=\rho^2(r_0)$, we then treat the 
 $|x|<r_0$ region as a point-like source with $\rho^2(|x|)-\rho^2(r_0)$. 
The radial boundary conditions can be simulated by modifying the DM density. 
First we note that sources located at a distance $r>R$ have a negligible effect, we then take $\rho=0$ when $r>R$. 
Second, radiation from  a source located near the boundary $r\leq R$ will be suppressed. 
To estimate this suppression we add {\it mirror} sources with negative charges on each side  of 
the boundary $r=R$. The long distance contribution from a point-like
source is proportionnal  to $K_0(r/a)\approx \exp(-r/a)$ where 
$a=1/\sqrt{k_0^2+k_c^2} \approx 2L/\pi$.  We then modify the DM density 
\begin{equation}
    \rho^2(r,z)  ~~~\to~~~  \Theta(R-r)\left(1-
\exp\left[-2(R -r)/a\right]\right) \rho^2(r,z) 
\end{equation} 
The suppression factor is significant only  in the region $R-L < r < R$.

The integrand in Eq.~(\ref{antiproton_answer}) (I(E)) is a smooth function that features only a slight energy dependence through
the terms $K(E)$ and $\Gamma_{tot}(E)$. To provide a fast 
calculation of the antiproton spectrum we interpolate the function $I(E)$ in the range $M_\chi - E_{\rm min}$, and multiply
this interpolated function by $f_{\bar{p}}(E)$ to obtain the final  result for all $E$. $E_{\rm min}$ is defined by the user. 
Note that points for interpolation are added  automatically  until the required precision is reached.

\subsubsection{Solar modulation}

When charged particles propagate through the solar system, they are affected by the solar wind and
lose energy. This effect leads to a shift in the energy distribution between the interstellar spectrum
and the spectrum at the  Earth, this  shift affects the low energy part of the spectrum.
We implement solar modulation using the force field approximation ~\cite{perko}.  In this approximation the
flux at the Earth is simply related to the flux at the heliospheric boundary ($\phi_h$),
\begin{equation}
\label{eq:solarmodulation}
\frac{d\Phi_\odot}{dE_\odot}=\frac{p^2_\odot}{p^2_h}\frac{d\Phi_h}{dE_h}
\end{equation}
where  $p_\odot$ and $p_h$ are the momenta at the Earth and the heliospheric boundary. The
energies at the two locations are simply related by
\begin{equation}
E_{\odot} = E_{IS} \, - \, \left| Z \right| \phi_{F} \;\; ,
\end{equation}
where the Fisk potential $\phi_F$ varies between 300MV  to 1000MV from the minimum to the maximum of
 solar activity.  The default value  is set to 500MV.

\section{Functions of micromegas}
\label{code}

A complete description of all  \micro~ functions is available in the online manual, 
 \verb|http://lapth.in2p3.fr/micromegas/|. This updated manual is also provided with the code, the file \verb|manual4.tex| can be found in the main
 \micro~2.4 directory. Here we describe the functions that
are specific to the indirect detection module.  A new feature of version 2.4 is the use of global parameters. A list of the
global parameters relevant for the indirect detection module and their default values 
 are given in Table 1. The numerical value for any of these parameters can be simply reset anywhere in the code.

\begin{table}[h]
\label{global}
\begin{center}
\begin{tabular}{|l|l|l|l|l|}
\hline
  Name      &Default value& Units &  Comments \\  \hline
Mcdm        &             &  GeV  & Mass of Dark Matter particle, $M_\chi$           \\ 
Rsun        & 8.          & kpc   & Distance from the Sun to the Galactic center, $r_{\odot}$\\
rhoDM       & 0.3         & $\rm{GeV/cm}^3$ & Dark Matter density at Rsun, $\rho_{\odot}$\\
Rdisk       & 20          & kpc   & Radius of the galactic diffusion disk, $R$ \\
K\_dif      & 0.0112      & $\rm{kpc^2/Myr}$ & Diffusion coefficient $K_0$\\
L\_dif      & 4           & kpc       & Half height of the galactic  diffusion zone $L$ \\
Delta\_dif  & 0.7         &           &Slope of diffusion coefficient, $\delta$\\ 
Tau\_dif    & $10^{16}$   &   s       &Positron energy loss time scale, $\tau_E$\\
Vc\_dif     & 12          &  km/s     &Convective velocity of Galactic vind , $V_C$\\
\hline
\end{tabular}
\caption{Global parameters of the indirect detection module}
\end{center}
\end{table}

\subsection{Spectra interpolation and display}

Various spectra of SM particles produced in DM annihilation processes are stored in
arrays containing  NZ=250 elements. The $i^{th}$ element of an array corresponds to 
 $dN/dz_i$ where $z_i=\log(E_i/M_\chi)$. Here \verb|E_i| is the kinetic
energy of the particle and $N$ stands for either the  
number of particles or a particle flux in $({\rm cm}^2 {\rm s} \;{\rm
sr})^{-1}$.\footnote{ Although not needed if one uses the interpolation functions, 
the value of $z_i$ can be obtained by the function 
$Zi(i)$. In the current version $Zi(i)=log(10^{-7})(i/NZ)^{1.5}$. }

The following functions can be used  for  interpolation and visualization \\ 

\noindent
$\bullet$  \verb|SpectdNdE(E,spectTab)|\\
interpolates the tabulated spectra  and returns the \verb|dN/dE| 
distribution where \verb|E| is the energy  in GeV.

\noindent
$\bullet$ \verb|zInterp(z,SpectTab)|\\
interpolates the tabulated spectra  and returns the \verb|dN/dz| distribution 
 where $z=\log(E/M_\chi)$,  here $z=0$ corresponds to $E=M_\chi$.

\noindent
$\bullet$ \verb|displaySpectrum(Spectrum,message,Emin,Emax,Units)|\\
displays the resulting spectrum, \verb|message| is a text string which gives a title to the  
graphic plot. \verb|Emin| and \verb|Emax| define energy cuts. If \verb|Units=0| the spectrum is written as a
function of $z$ otherwise the spectrum is a function of the energy in GeV.

\subsection{Annihilation spectra}
$\bullet$ \verb|calcSpectrum(key,Sg,Se,Sp,Sne,Snm,Snl,&err)|\\
calculates  the spectra  of DM annihilation 
at rest and returns $\sigma v$ in $cm^3/s$ . The calculated spectra
for $\gamma$, $e^+$, $\bar{p}$, $\nu_e$, $\nu_{\mu}$, $\nu_{\tau}$ 
are stored in arrays of dimension \verb|NZ| as described above: \verb|Sg|, \verb|Se|, \verb|Sp|, 
\verb|Sg|, \verb|Sne|, \verb|Snm|, \verb|Snl|. 
 To remove the calculation of a given spectra, substitute  
\verb|NULL| for the corresponding argument. 
\verb|key| is a switch to include polarisation of W,Z bosons (\verb|key=1|) or
 photon radiation (\verb|key=2|).  Photon radiation is added to all subprocesses when computing the photon spectrum while
only the 3-body process $\chi\chi\rightarrow e^+e^-\gamma$ is included  for the positron spectrum. 
When \verb|key=4| the cross sections for each annihilation channel are written on the screen. More than one option
can be switched on simultaneously by adding the corresponding values for \verb|key|. 
For example both the W polarisation and photon radiation effects  are included if
\verb|key=3|.
A problem in the spectrum calculation will produce a non zero error code, $err\neq 0$. 
  
\noindent
$\bullet$ \verb|spectrInfo(Xmin,spectrTab,&Ntot,&Xtot)|\\
provides information on the spectra generated. Here \verb|Xmin| defines the minimum 
cut for the energy fraction x=E/Mcdm, \verb|Ntot| and \verb|Xtot| are calculated parameters 
which give on average the total number and the energy fraction of the final particles 
produced per collision. Note that the upper limit is Xtot=2.

\noindent
$\bullet$ \verb|basicSpectra(pdgN,outN,Spectr)|\\
is a routine for model independent studies that computes the spectra of outgoing particles as obtained by PYTHIA and 
writes the result in an array of dimension NZ, \verb|Spectr|.
\verb|pdgN| is the PDG code of the particles produced in the annihilation of a pair of 
dark matter particles. The spectrum for polarised W's or Z's is obtained by substituting  \verb|pdgN+'T'| (transverse) or
\verb|pdgN+'L'| (longitudinal) for the PDG code of the W(Z), \verb|pdgN|=24(23).
\verb|outN|  specifies the outgoing particle,
$$ {\rm outN} = \{0,1,2,3,4,5\} \;\; {\rm for}\;\; \{\gamma,   e^+,  p^-, \nu_e,
\nu_{\mu},\nu_{\tau}\} $$
Note that the  propagation routines for $e^+,\bar{p},\gamma$ can be used after this routine as usual.

\noi$\bullet$ \verb|loopGamma(&vcs_gz,&vcs_gg)|\\
calculates $\sigma v$ in ${cm^3}/{s}$ for the loop induced  neutralino
annihilation into $\gamma Z$ and  $\gamma \gamma$. In case of problem the function returns a non-zero value. 
This function is available only for the MSSM.

\subsection{Distribution of Dark Matter in the Galaxy.}
To compute the signal from an indirect detection experiment one has 
to take into account the dark matter distribution in the Galaxy. Both 
the DM density profile as well as the clump profile,  Eq.~\ref{halo}, have to be defined.
The DM density at the Sun, $\rho_\odot$ as well as  $r_\odot$, the distance from the center of the Galaxy to the Sun
 are defined  by the global variables \verb|rhoDM| and \verb|Rsun|.

\noindent
$\bullet$ \verb|setHaloProfiles(|$F_{halo},F_{clump}$\verb|)|\\
allows to change both the halo and the clump profile. Any sphericaly symmetric DM halo
profile can be defined.

\noindent
$\bullet$ \verb|hProfileABG(r)|\\
is the default halo  density profile, Eq.~\ref{halo}

\noindent
$\bullet$ \verb|setProfileABG(alpha,beta,gamma,a)|\\
resets the parameters of the  DM distribution profile. The default
parameters correspond to the NFW profile, $\alpha=1,\beta=3,\gamma=1,a=20[kpc]$.

\noindent
$\bullet$ \verb|hProfileEinasto(r)|\\
is the Einasto halo density profile, eq.~\ref{einasto}.

\noindent
$\bullet$\verb|setProfileEinasto(alpha)| \\
sets the parameter $\alpha$  for the  Einasto profile, the  default value is $\alpha=0.17$.

\noindent
$\bullet$ \verb|noClumps(r)|\\
is a non clumpy profile which is used by default. It returns $1$  for any argument.

\subsection{Particle  propagation.}

The spectrum of charged particles observed strongly depends on their propagation
in the Galactic Halo. The propagation depends on the global parameters 
\begin{verbatim}
       K_dif, Delta_dif, L_dif, Rsun, Rdisk
\end{verbatim}
as well as 
\begin{verbatim}
 Tau_dif (positrons), Vc_dif (antiprotons)
\end{verbatim}

\noindent  
$\bullet$ \verb|posiFluxTab(Emin,sigmav, Se,  Sobs)|\\
computes the positron flux at the Earth. Here \verb|sigmav| and \verb|Se| are values obtained by 
\verb|calcSpectrum|.  \verb|Sobs| is the positron spectrum after propagation. \verb|Emin| is the energy cut to be defined by the user. Note that
a low value for \verb|Emin| increases the computation time.
The  format is the same as for the initial spectrum. The function  
\verb|SpectrdNdE(E,Sobs)| described above can also be used for interpolation, in this case the flux
returned in (GeV s ${\rm cm}^2 {\rm sr})^{-1}$). 

\noindent
$\bullet$ \verb|pbarFlux(E,dSigmavdE)|\\
computes the antiproton flux for a given energy {\tt E} and a 
differential cross section for antiproton production, {\tt dSigmavdE}.
For example, one can substitute\\ {\tt dSigmavdE}=$\sigma v${\tt
SpectdNdE(E,SpP)} \\
where $ \sigma v$ and {SpP} are obtained by {\tt calcSpectrum}.
This function does not depend on the details of the particle physics model and allows to analyse the dependence on the
parameters of the propagation model.

\noindent
$\bullet$ \verb|pbarFluxTab(Emin,sigmav, Sp,  Sobs)|\\
computes the antiproton flux, this function works like \verb|posiFluxTab|,

\noindent
$\bullet$ \verb|solarModulation(Phi, mass, stellarTab, earthTab)|\\
takes into account modification of the interstellar positron/antiproton flux 
caused by the electro-magnetic fields in the solar system. Here \verb|Phi| is the
effective Fisk potential in MeV, \verb|mass| is the particle mass,
\verb|stellarTab| describes the interstellar flux, \verb|earthTab| 
is the calculated particle flux in the Earth orbit.

The photon  flux does not depend on the  diffusion model parameters but on the angle
$\phi$ between the line of sight and the center of the galaxy as well as on the annihilation spectrum
into photons

\noindent
$\bullet$ \verb|gammaFluxTab(fi,dfi,sigmav,Sg,Sobs)|\\
multiplies the annihilation photon spectrum  with the integral over the line of sight
and over the opening angle to give the photon flux. 
\verb|fi| is the angle between the line of sight and the center of the
galaxy,   \verb|dfi| is half the cone angle which characterizes the detector resolution
(the solid angle is  $2\pi (1-cos(dfi)$) ,  
 \verb|sigmav| is the annihilation cross section, \verb|Sg| is the DM annihilation spectra.
\verb|Sobs| is the spectra observed.
Note that  \verb|Emin| is  not specified, since this function is not time 
consuming the integration is done for all energies. 

The function \verb|gammaFluxTab| can be used  for the neutrino spectra as well.

\noindent
$\bullet$ \verb|gammaFlux(fi,dfi,vcs)|\\
is the same function as \verb|gammaFluxTab| above  but corresponds to  
a discrete photon spectrum. \verb|vcs| is the annihilation cross section, for instance in the {\t MSSM} it is 
calculated with the \verb|loopGamma| function. The function
returns the number of photons per $cm^2$ of detector surface
 per second. Note that for $\chi\chi \to \gamma\gamma$ the  
result should be multiplied by a factor $2$ as each annihilation leads to the 
production of two photons. 

Note that for \verb|solarModulation| and for  all \verb|*FluxTab| 
routines one can use  the same array for the spectrum before and after propagation.

\section{Examples and results}
\label{examples}

\subsection{Sample output}

Running the main.c sample file in the \micro/MSSM directory choosing the options 
\verb|MASSES_INFO,CONSTRAINTS,OMEGA|, \verb|INDIRECT_DETECTION|, \verb|CDM_NUCLEON|  will 
lead to the following output. 

\small{
\begin{verbatim}

========= SLHA file input =========
Initial file  "model2.slha"
Warnings from spectrum calculator:
Model: model2

Dark matter candidate is '~o1' with spin=1/2  mass=1.48E+02

~o1 = 0.833*bino -0.114*wino -0.448*higgsino1 -0.303*higgsino2

=== MASSES OF HIGGS AND SUSY PARTICLES: ===
Higgs masses and widths
    h   123.08 2.59E-03
    H  1000.20 1.19E+01
   H3  1000.00 1.20E+01
   H+   998.63 1.18E+01

Masses of odd sector Particles:
~o1  : MNE1  =   147.7 || ~2+  : MC2   =   189.9 || ~o2  : MNE2  =   198.2 
~o3  : MNE3  =   211.1 || ~o4  : MNE4  =   345.1 || ~1+  : MC1   =   345.3 
~g   : MSG   =  1108.9 || ~t1  : MSt1  =  2418.8 || ~b1  : MSb1  =  2496.4 
~ne  : MSne  =  2499.2 || ~nm  : MSnm  =  2499.2 || ~nl  : MSnl  =  2499.2 
~uL  : MSuL  =  2499.4 || ~cL  : MScL  =  2499.4 || ~uR  : MSuR  =  2499.8 
~cR  : MScR  =  2499.8 || ~l1  : MSl1  =  2499.8 || ~sL  : MSsL  =  2500.1 
~dL  : MSdL  =  2500.1 || ~mL  : MSmL  =  2500.3 || ~eL  : MSeL  =  2500.4 
~eR  : MSeR  =  2500.4 || ~mR  : MSmR  =  2500.5 || ~dR  : MSdR  =  2500.7 
~sR  : MSsR  =  2500.7 || ~l2  : MSl2  =  2501.0 || ~b2  : MSb2  =  2504.4 
~t2  : MSt2  =  2589.4 || 


==== Physical Constraints: =====
deltartho=5.70E-06
gmuon=-8.31E-11
bsgnlo=3.94E-04
bsmumu=3.02E-09
btaunu=9.96E-01
MassLimits OK

==== Calculation of relic density =====
Xf=2.45e+01 Omega=1.13e-01

Channels which contribute to 1/(omega) more than 1%.
Relative contrubutions in % are displyed
 60% ~o1 ~o1 -> W+ W- 
 26% ~o1 ~o1 -> Z Z 
  9% ~o1 ~o1 -> Z h 
  1% ~o1 ~o1 -> h h 
The rest 3.28 %

==== Indirect detection =======
    Channel          vcs[cs^3/s]
  ~o1,~o1 -> h h        1.30E-43
  ~o1,~o1 -> Z h        1.31E-27
  ~o1,~o1 -> Z Z        5.03E-27
  ~o1,~o1 -> W+ W-      1.20E-26
sigmav=1.83E-26[cm^3/s]
Photon flux  for angle of sight f=0.00[rad]
and spherical region described by cone with angle 0.04[rad]
Photon flux = 2.01E-14[cm^2 s GeV]^{-1} for E=73.8[GeV]
Gamma  ray lines:
E=1.34E+02[GeV]  vcs(Z,A)= 1.37E-29[cm^3/s], flux=7.19E-14[cm^2 s]^{-1}
E=1.48E+02[GeV]  vcs(A,A)= 2.47E-30[cm^3/s], flux=2.58E-14[cm^2 s]^{-1}
N_=19
Positron flux  =  8.09E-12[cm^2 sr s GeV]^{-1} for E=73.8[GeV] 
N_grid=9
Antiproton flux  =  1.77E-11[cm^2 sr s GeV]^{-1} for E=73.8[GeV] 

==== Calculation of CDM-nucleons amplitudes  =====
CDM-nucleon micrOMEGAs amplitudes:
proton:  SI  -4.570E-09  SD  -6.116E-07
neutron: SI  -4.582E-09  SD  5.355E-07
CDM-nucleon cross sections[pb]:
 proton  SI 9.015E-09  SD 4.844E-04
 neutron SI 9.061E-09  SD 3.714E-04
\end{verbatim}
}
 
\normalsize
\subsection{Comparison with other packages}

We have performed many tests to check the consistency of the results obtained with \micro. 
 For the treatment of the hadronization process, we have computed  the fragmentation functions
 with \py (those are implemented in \micro) and with \HE. 
Despite the fact that the two codes are based on 
 different modelling of the hadronization process, the 
functions $dN/dE$ obtained   with \py~  and \HE~  are very similar. 
A comparison  between the two codes  in the case of a 500 GeV  annihilating into $b\bar{b}$ pairs 
 has been performed and  gave similar results.

\begin{table}[bh]
\centering
	\begin{tabular}{c|c|c|c|c}
	 &  \multicolumn{2}{c|}{Model 1} & \multicolumn{2}{c}{Model 2}  \\ \hline
	 &  \multicolumn{2}{c|}{$\mu=-440, M_A=1000$} & \multicolumn{2}{c}{$\mu=-200, M_A=1000$}  \\ \hline
	 &  \multicolumn{2}{c|}{$M_2=800, M_0=2500$} & \multicolumn{2}{c}{$M_2=320,M_0=2500$}  \\ \hline
	 &  \multicolumn{2}{c|}{$A_t=A_b=1000$} & \multicolumn{2}{c}{$A_t=-A_b=-2500$}  \\ \hline
	  &  \multicolumn{2}{c|}{$\tan\beta=10$} & \multicolumn{2}{c}{$\tan\beta=8.$}  \\ \hline
	 &  \micro & \ds & \micro & \ds \\ \hline
	 $M_\chi$ (GeV)& 386.7 & 386.7 & 147.7 & 147.7\\ \hline
	 $\sigma v  \rm (cm^3s^{-1})$ & $1.97\times 10^{-26}$ & $2.32\times 10^{-26}$ & $1.83\times 10^{-26}$ & $1.97 \times 10^{-26}$\\ \hline
	 Main final states 	& $t\bar{t}$ (77.7\%) & $t\bar{t}$ (80.4\%) & $W^+W^-$ (65.6\%) & $W^+W^-$ (65.2\%)\\
	 				& $W^+W^-$ (8.9\%) & $W^+W^-$ (8.1\%) & $Z^0Z^0$ (27.5\%)& $Z^0Z^0$ (26.8\%)\\ 
					& $ZZ$ (6.5\%) & $ZZ$ (5.8\%) & $Z^0 h$ (7.2\%)& $Z^0 h$ (7.2\%)\\
	\end{tabular}
	\caption{\label{summarytable} Main features of the two sets of parameters used for the comparison}
\end{table}

For the signal itself, we compare the results obtained with \ds~5.0.4 and \micro2.4. 
To perform the comparison, we generate MSSM points with \ds~ and used the resulting SLHA file~\cite{SLHA}
as an input for \micro~. 
Each  package computes the annihilation cross-section, 
branching ratios, propagation and distribution. We have chosen two sample models in the MSSM, the input parameters at
the weak scale are specified in Tab.~\ref{summarytable} where all dimensionful parameters are in GeV.
We assume  $2M_1=M_2=M_3/3$ as would be obtained in a GUT model with gaugino mass universality, a common
sfermion mass at the weak scale, $m_0$ and vanishing trilinear couplings of sleptons and of the first and second generations of
squarks. The files used for this comparison are provided in the MSSM directory of ~\micro 
(\verb|model1.slha| and \verb|model2.slha|).  The main features of the sample models are 
summarized in Tab.~\ref{summarytable}. The first obsevation is that the total annihilation cross section
as well as the individual channel contribution can vary by up to 17\%. This is particularly important if the main
annihilation channels are into third generation quarks (Model 1). Indeed \ds~ and \micro~ use a different prescription for the
running of quark Yukawa couplings. 
Part of the difference between the two codes is due to a different prescription for $\sin^2\theta_W$, 
an On-Shell scheme in \ds~ and $\overline{MS}$ in \micro~.

\begin{figure}[h]
\centering
\includegraphics[width=14cm]{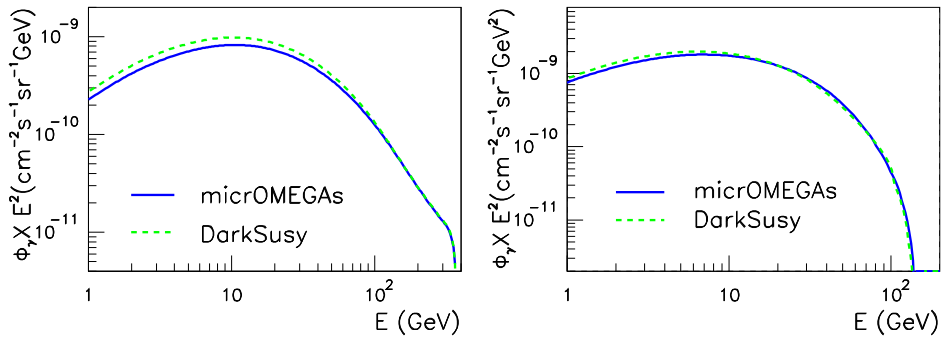}
\vspace{-1cm}
\caption{\label{comparison_gamma}$\gamma$-ray signal for Model 1(left) and Model 2(right) with \micro~  and \ds}
\end{figure}

To compare the $\gamma$-ray spectrum,  we use an NFW profile and compute 
the signal from the Galactic center in a cone of $1^o$ opening angle (corresponding to a solid angle of 
$\Delta \Omega = 10^{-3}\;\rm sr$). Fig.~\ref{comparison_gamma} displays the results 
for both models. For this we have switched on the gauge boson polarisation and included 
the contribution from 3-body final states with a photon.  The small difference between the
two spectra is explained by the difference in $\sigma v$.
Notice that on these plots, the flux is $E^2$-corrected.

For the one-loop processes leading to a monochromatic gamma-ray line, 
a detailed comparison between \ds~ and SloopS was performed in ~\cite{Boudjema:2005hb}.  The two codes agreed very well for the  annihilation of a  neutralino pair into two photons. We also find good agreement ($<5\%$) 
between  \micro~ and \ds~ for our  two test models. On the other hand in ~\cite{Boudjema:2005hb} it was pointed out that large differences  between 
\ds~ and {\tt SloopS} can  occur for the $\gamma Z$ one-loop process ($>30\%$). This is mainly because some diagrams were neglected in  \ds~, these give a non negligible contribution for the case of a higgsino LSP.

\begin{figure}[h]
\centering
\includegraphics[width=14cm]{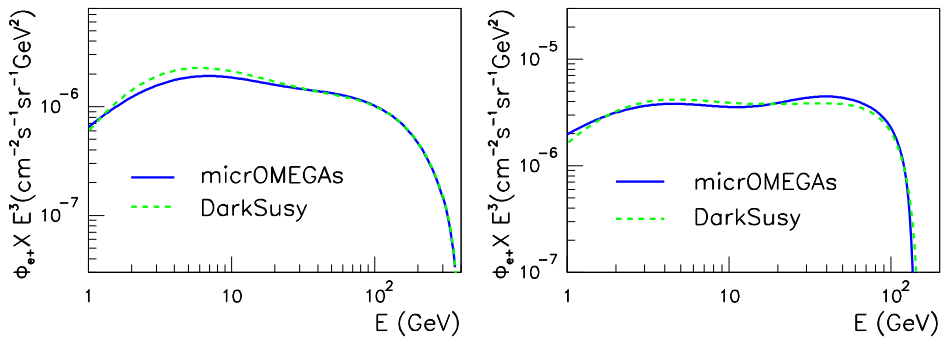}
\vspace{-1cm}
\caption{\label{comparison_posit} Positron signal for Model 1(left) and Model 2(right)
 with \micro~ and \ds. Here we have set $\delta=0.6$, $K_0=0.03607~{\rm kpc}^2/{\rm Myr}$, $L=4$~kpc and
 $V_C=10$~km/s.}
\end{figure}

\begin{figure}[h]
\centering
\includegraphics[width=14cm]{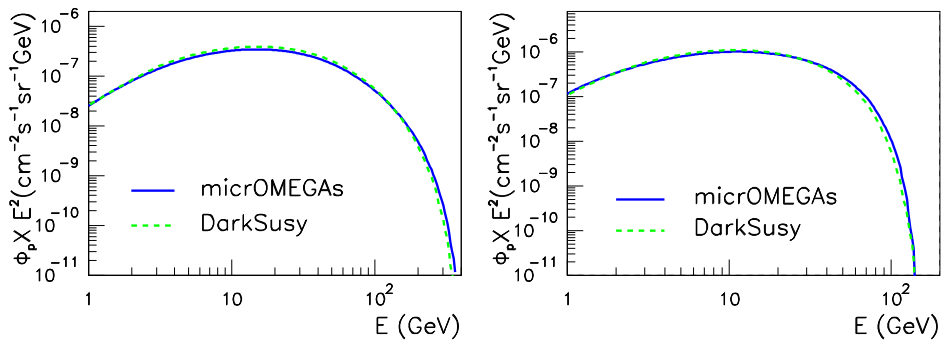}
\caption{\label{comparison_pbar} Antiproton signal for Model 1(left) and Model 2(right)
 with \micro~ and \ds.  Same propagation parameters as above. Solar modulation is included with 
 $\phi_F=320$~MV.}
\end{figure}

In the case of positrons, the comparison is shown on Fig.~\ref{comparison_posit}. The 
propagation parameters  are the default parameters of \ds~,   $R=30 {\rm kpc}$, $h=0.1
{\rm kpc}$, $L= 4{\rm kpc}$, $K_0=0.0826 {\rm kpc^2/Myr}$, $\delta=0.6$, $V_C=10{\rm km/s}$.
Note that the space diffusion coefficient $K$
 is defined at a reference energy of $4 {\rm GeV}$ in \ds~ rather than 1GeV in ~\micro~. To reproduce similar propagation parameters 
 one must therefore rescale the coefficient $K_0$ in ~\micro. Furthermore  for positrons \ds~ assumes that $\delta=0$ below 
 $p=4{\rm GeV}$. The difference between the two codes is small (few percent) once taking into account the correction due to 
 the initial $\sigma v$.  Larger differences are
 found at low energies and could be due to the different treatment 
of the positron propagation in that energy range. However, as shown below in section 5.3, 
the observed differences are much below the theoretical uncertainty induced by the propagation parameters.
Good agreement between the two codes is also recovered for the antiproton signal, see 
Fig.~\ref{comparison_pbar}. 

\subsection{Theoretical uncertainties related to propagation}

\begin{figure}[h]
\centering
\includegraphics[width=14cm]{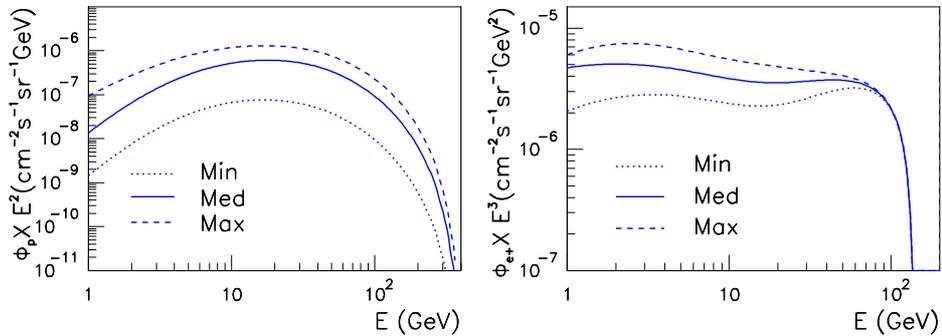}
\vspace{-1cm}
\caption{\label{propagation}Propagation related uncertainty for antiprotons  in Model 1(left) and for positrons in Model 2 (right).
The propagation parameters are listed in Table 2.}
\end{figure}

The way the propagation of charged particles is handled in the code allows to very quickly estimate the propagation 
related uncertainty. 
For this we compare antiprotons and positrons signals using extreme values for the propagation parameters  
that are still  allowed by all cosmic ray measurements.  
 The curves labelled "Min", "Med", "Max" in Fig.~\ref{propagation} correspond to the parameters specified in Tab.~\ref{tab:BC}.
Recall that these parameters were selected to reproduce the data on the B/C ratio~\cite{Maurin:2001sj,Donato:2003xg}. 
For antiprotons the uncertainty exceeds one order of magnitude over the full range of energy while for positrons
the uncertainty is large mainly at low energies. Actually high energy positrons do not suffer from propagation uncertainties since they are
produced locally. 

\section{Conclusions}
With the new module for indirect detection presented here, \micro~ is a comprehensive code for 
the study of the properties of
weakly interacting cold dark matter candidates in various extensions of the standard model. 
The main features of this new version include  the propagation of charged cosmic rays
and some higher order processes in DM annihilation. In particular the code contains a full computation of photon
radiation.
These new features are available for any model with a weakly interacting particle that is already implemented in \micro.
In addition  a specific code  for the computation of the loop-induced gamma-ray line in the MSSM is included. 
The modular and flexible structure of the code makes it simple for the user to improve
some specific part of the code, for example adding  new
functions that take into account the presence of dark matter clumps. 
The main observable
that is not yet included in the code is the neutrino signal associated with dark matter capture in the Sun or in
the Earth. This feature will be available in the next upgrade of \micro. 
The code is available at http://lapth.in2p3.fr/micromegas/.

\section*{Acknowledgements}

We thank G. Chalons, G. Drieu La Rochelle, J. da Silva, S. Kulkarni, S. Kraml and Y. Mambrini for
their help in debugging the code. 
This work was  supported in part by the GDRI-ACPP of CNRS and by the ANR project {\tt ToolsDMColl}, BLAN07-2-194882.
This work  was  also supported by the Russian foundation for Basic Research, grant
RFBR-08-02-00856-a,  RFBR-08-02-92499-a, RPBR-10-02-01443-a and by a State contract No.02.740.11.0244.

\appendix

\end{document}